\documentclass[letterpaper, 10pt, conference]{ieeeconf}      

\IEEEoverridecommandlockouts                              
\usepackage{amsmath}    
\usepackage{amsfonts}
\usepackage{graphicx}   
\usepackage{subcaption}
\usepackage{epsfig} 
\usepackage{cancel}
\usepackage{amssymb}
\usepackage{color}
\usepackage{my_macros}
\usepackage[ruled,vlined,titlenotnumbered]{algorithm2e} 

\title{\LARGE \bf
Safe Sequential Path Planning of Multi-Vehicle Systems via Double-Obstacle Hamilton-Jacobi-Isaacs Variational Inequality}

\author{Mo Chen, Jaime F. Fisac, Shankar Sastry, Claire J. Tomlin
\thanks{This work has been supported in part by NSF under CPS:ActionWebs (CNS-931843), by ONR under the HUNT (N0014-08-0696) and SMARTS (N00014-09-1-1051) MURIs and by grant N00014-12-1-0609, by AFOSR under the CHASE MURI (FA9550-10-1-0567). The research of J.F. Fisac has received funding from the ``la Caixa" Foundation.}
\thanks{All authors are with the Department of Electrical Engineering and Computer Sciences, University of California, Berkeley. \{mochen72, jfisac, sastry, tomlin\}@eecs.berkeley.edu}
}

\begin{document}
\maketitle
\thispagestyle{empty}
\pagestyle{empty}

\begin{abstract}
We consider the problem of planning trajectories for a group of $N$ vehicles, each aiming to reach its own target set while avoiding danger zones of other vehicles. The analysis of problems like this is extremely important practically, especially given the growing interest in utilizing unmanned aircraft systems for civil purposes. The direct solution of this problem by solving a single-obstacle Hamilton-Jacobi-Isaacs (HJI) variational inequality (VI) is numerically intractable due to the exponential scaling of computation complexity with problem dimensionality. Furthermore, the single-obstacle HJI VI cannot directly handle situations in which vehicles do not have a common scheduled arrival time. Instead, we perform sequential path planning  by considering vehicles in order of priority, modeling higher-priority vehicles as time-varying obstacles for lower-priority vehicles. To do this, we solve a double-obstacle HJI VI which allows us to obtain the reach-avoid set, defined as the set of states from which a vehicle can reach its target while staying within a time-varying state constraint set. From the solution of the double-obstacle HJI VI, we can also extract the latest start time and the optimal control for each vehicle. This is a first application of the double-obstacle HJI VI which can handle systems with time-varying dynamics, target sets, and state constraint sets, and results in computation complexity that scales linearly, as opposed to exponentially, with the number of vehicles in consideration.
\end{abstract}

\section{Introduction}
Consider a group of autonomous vehicles trying to perform a task or reach a goal which may be time-varying in their joint state space, while avoiding obstacles and other vehicles. Providing safety and performance guarantees for such a multi-agent autonomous system (MAAS) is very relevant practically: Recently, there has been a growing interest in using unmanned aerial vehicles (UAVs) for civil applications, as companies like Amazon and Google are looking in the near future to send UAVs into the airspace to deliver packages \cite{primeAir,projectWing}. Government agencies such as the Federal Aviation Administration (FAA) and National Aeronautics and Space Administration (NASA) of the United States are also expressing growing interest in analyzing these problems in order to prevent airspace conflicts that could arise with the introduction of potentially many UAVs in an urban environment \cite{faa13}. In addition, UAVs can be used not only to deliver packages quickly, but in any situation where fast response is desired. For example, UAVs can provide emergency supplies to disaster-struck areas that are otherwise difficult to reach \cite{debusk10}.

In general, MAASs are difficult to analyze due to their inherent high dimensionality. 
 MAASs also often involve aspects of cooperation and asymmetric goals among the vehicles or teams of vehicles, making their analysis particularly interesting. 
MAASs have been explored extensively in the literature. Some researchers have done work on multi-vehicle path planning in the presence of other unknown vehicles or moving entities with assumptions on their specific control strategies \cite{chasparis05}. In a number of formulations for safe multi-vehicle navigation, these assumed strategies induce velocity obstacles that vehicles must avoid to maintain safety \cite{fiorini98, vandenberg08}. Researchers have also used potential functions to perform collision avoidance while maintaining formation given a predefined trajectory \cite{saber02,chuang07}. However, these bodies of work have not considered trajectory planning and collision avoidance simultaneously.

One well-known technique for optimal trajectory planning under disturbances or adversaries is reachability analysis, in which one computes the reach-avoid set, defined as the set of states from which the system can reach a target set while remaining within a state constraint set for all time. For reachability of systems of up to five dimensions, single-obstacle Hamilton-Jacobi-Isaacs (HJI) variational inequalities (VI) \cite{mitchell05,bokanowski10} have been used in situations where obstacles and target sets are static. Another HJI VI formulation \cite{barron89} is able to handle problems with moving target sets with no obstacles. 

A major practical appeal of the above approaches stems from the availability of modern numerical tools such as \cite{mitchell05, sethian96, osher02, LSToolbox}, which can efficiently solve HJI equations when the problem dimension is low. These numerical tools have been successfully used to solve a variety of differential games, path planning problems, and optimal control problems\cite{mitchell05,ding08,huang11}. 
 Despite the power of the previous HJI formulations, the approaches become numerically intractable very quickly as the number of vehicles in the system is increased. This is because the numerical computations are done on a grid in the joint state space of the system, resulting in an exponential scaling of computation complexity with respect to the dimensionality of the problem. Furthermore, state constraint sets, while useful for modeling unsafe vehicle configurations, are required to be time-invariant in \cite{mitchell05, bokanowski10, mitchell-thesis}. To solve problems involving time-varying state constraints, \cite{bokanowski11} proposed to augment the state space with time; however, this process introduces an extra state space dimension, resulting in added computation complexity.

Recently, \cite{fisac15} presented a double-obstacle HJI VI which handles problems in which the dynamics, target sets, and state constraint sets are all time-varying, and provided a numerical implementation based on well-known schemes. The formulation does not introduce any additional computation overhead compared to the above-mentioned techniques, yet it still maintains the same guarantees on the system's safety and performance. In this paper, we provide a first application of the theory presented in \cite{fisac15}. As a point of clarification, ``obstacles" in the context of HJI VIs refer to the effective constraints in the HJI VI, while obstacles in the state space represent physical obstacles that vehicles must avoid.

Our contributions are as follows. First, we formulate a multi-vehicle collision avoidance problem involving $N$ autonomous vehicles. Each vehicle seeks to get to its own target sets while avoiding obstacles and collision with all other vehicles. To reduce the problem complexity to make the problem tractable, we assign a priority to each vehicle, and model higher-priority vehicles as time-varying obstacles that need to be avoided. We then utilize the double-obstacle HJI VI proposed in \cite{fisac15} to compute reach-avoid sets to plan trajectories for vehicles in order of priority. This way, we are able to offer a tractable solution that scales linearly, as opposed to exponentially, with the number of vehicles. We demonstrate the scalability of our approach in a four-vehicle system.

\section{Problem Formulation \label{sec:formulation}}
Consider $N$ vehicles $P_i,i=1\ldots,N$, each trying to reach one of $N$ target sets $\target_i,i=1\ldots,N$, while avoiding obstacles and collision with each other. Each vehicle $i$ has states $\x_i\in \R^{n_i}$ and travels on a domain $\amb=\obs \cup \free\in\R^p$, where $\obs$ represents the obstacles that each vehicle must avoid, and $\free$ represents all other states in the domain on which vehicles can move. Each vehicle $i = 1,2,\ldots,N$ moves with the following dynamics for $t\in[\tnow_i, \tf_i]$:

\begin{equation} \label{eq:dyn}
\dotx_i = f_i (t, \x_i, \ctrl_i), \quad\x_i(\ti_i) = \x_i^0 
\end{equation}

\noindent where $\x_i^0$ represents the initial condition of vehicle $i$, and $\ctrl_i(\cdot)$ represents the control function of vehicle $i$. In general, $f_i(\cdot,\cdot,\cdot)$ depends on the specific dynamic model of vehicle $i$, and need not be of the same form across different vehicles. Denote $\pos_i\in\R^p$ the subset of the states that represent the position of the vehicle. Given $\pos_i^0\in\free$, we define the admissible control function set for $P_i$ to be the set of all control functions such that $\pos_i(t) \in \free \forall t\ge \ti_i$. Denote the joint state space of all vehicles $\x \in \R^n$ where $n = \sum_i n_i$, and their joint control $\ctrl$.

We assume that the control functions $\ctrl_i(\cdot)$ are drawn from the set $\ctrlf_i := \{\ctrl_i: [\tnow_i, \tf_i] \rightarrow \ctrlin_i, \text{measurable}$\footnote{
A function $f:X\to Y$ between two measurable spaces $(X,\Sigma_X)$ and $(Y,\Sigma_Y)$ is said to be measurable if the preimage of a measurable set in $Y$ is a measurable set in $X$, that is: $\forall V\in\Sigma_Y, f^{-1}(V)\in\Sigma_X$, with $\Sigma_X,\Sigma_Y$ $\sigma$-algebras on $X$,$Y$.}\} where $\ctrlin_i \in \R^{n^\ctrl_i}$ is the set of allowed control inputs. Furthermore, we assume $f_i(t,\x_i, \ctrl_i)$ is bounded, Lipschitz continuous in $\x_i$ for any fixed $t,\ctrl_i$, and measurable in $t, \ctrl_i$ for each $\x_i$. Therefore given any initial state $\x_i^0$ and any control function $\ctrl_i(\cdot)$, there exists a unique, continuous trajectory $\x_i(\cdot)$ solving (\ref{eq:dyn}) \cite{coddington55}.

The goal of each vehicle $i$ is to arrive at $\target_i \subset \R^{n_i}$ at or before some scheduled time of arrival (STA) $\tf_i$ in minimum time, while avoiding obstacles and danger with all other vehicles. The target sets $\target_i$ can be used to represent desired kinematic quantities such as position and velocity and, in the case of non-holonomic systems, quantities such as heading angle.  $\tnow_i$ can be interpreted as the earliest start time (EST) of vehicle $i$, before which the vehicle may not depart from its initial state. Further, we define $\ti_i$, the latest (acceptable) start time (LST) for vehicle $i$. Our problem can now be thought of as determining the LST $\ti_i$ for each vehicle to get to $\target_i$ at or before the STA $\tf_i$, and finding a control to do this safely. If the LST is before the EST $\ti_i < \tnow_i$, then it is infeasible for vehicle $i$ to arrive at $\target_i$ at or before the STA $\tf_i$. Comparing $\ti_i$ and $\tnow_i$ is feasibility problem that may arise in practice; however, for simplicity of presentation, we will assume that $\tnow_i\le \ti_i \forall i$.

\begin{figure}
	\centering
	\includegraphics[width=0.35\textwidth]{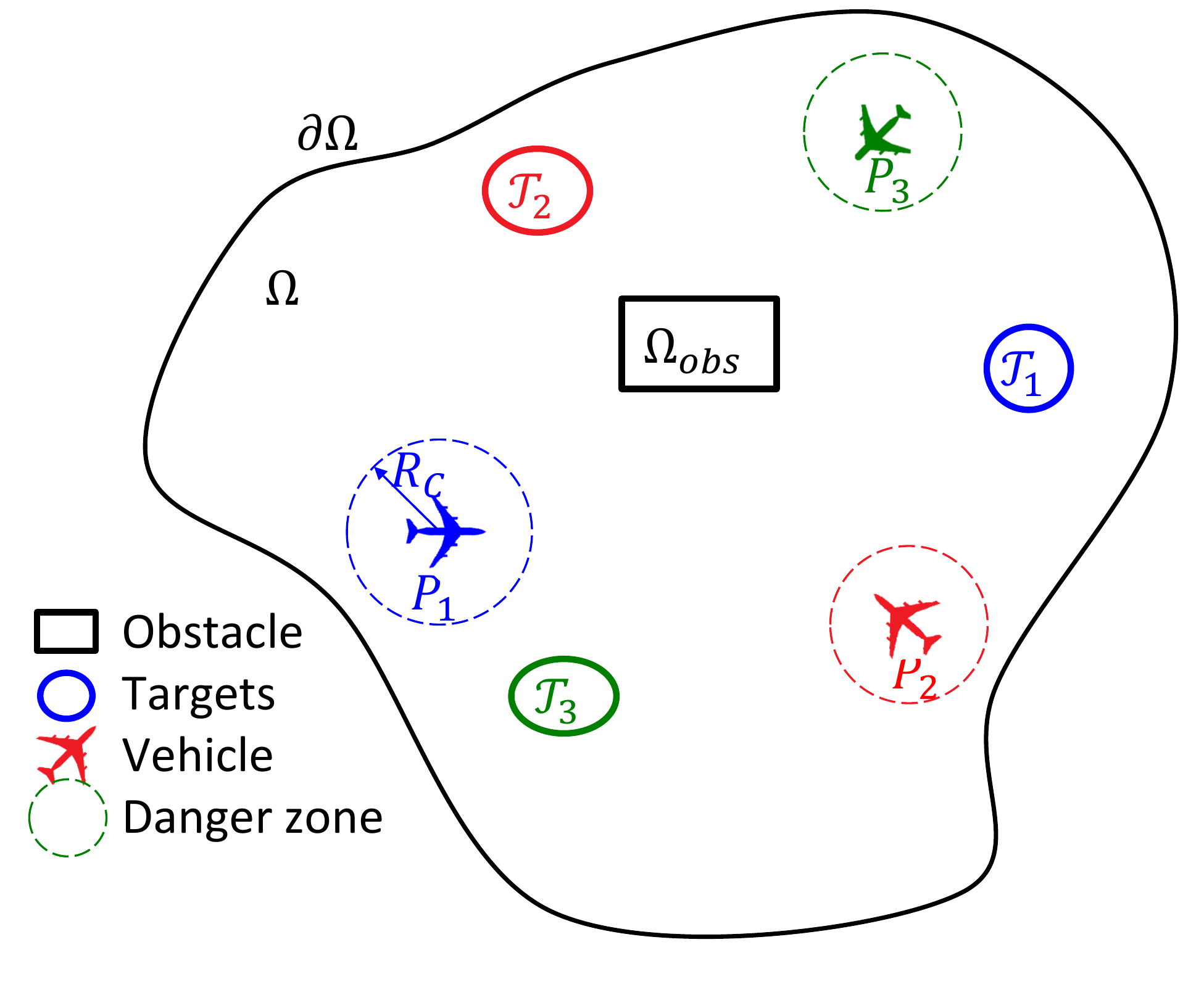}
	\caption{An illustration of the problem formulation with three vehicles. Each vehicle $P_i$ seeks to reach its target set $\target_i$ by time $t=\tf_i$, while avoiding physical obstacles $\obs$ and the danger zones of other vehicles.}
	\label{fig:formulation}
\end{figure}

Danger is described by sets $\danger_{ij}(\x_j) \subset \amb$. In general, the definition of $\danger_{ij}$ depends on the conditions under which vehicles $i$ and $j$ are considered to be in an unsafe configuration, given the state of vehicle $j$. Here, we define danger to be the situation in which the two vehicles come within a certain radius $\Rc$ of each other: $\danger_{ij}(\x_j) = \{\x_i: \| \pos_i - \pos_j\|_2 \le \Rc \}$. Such a danger zone is also used by the FAA \cite{paglione99}. An illustration of the problem setup is shown in Figure \ref{fig:formulation}.

In general, the above problem must be analyzed in the joint state space of all vehicles, making the solution intractable. In this paper, we will instead consider the problem of performing path planning of the vehicles in a sequential manner. Without loss of generality, we consider the problem of first fixing $i=1$ and determining the optimal control for vehicle $1$, the vehicle with the highest priority. The resulting optimal control $\ctrl_1$ sends vehicle $1$ to $\target_1$ in minimum time. 

Then, we plan the minimum time trajectory for each of the vehicles $2,\ldots,N$, in decreasing order of priority, given the previously-determined trajectories for higher-priority vehicles $1,\ldots,i-1$. We assume that all vehicles have complete information about the states and trajectories of higher-priority vehicles, and that all vehicles adhere to their planned trajectories. Thus, in planning its trajectory, vehicle $i$ treats higher-priority vehicles as known time-varying obstacles. 

With the above sequential path planning (SPP) protocol and assumptions, our problem now reduces to the following for vehicle $i$. Given $\x_j(\cdot), j=1,\ldots,i-1$, determine $\ctrl_i(\cdot)$ that maximizes $\ti_i$ and such that $x_i(\tau) \in \target_i, \tau\le \tf_i$.

\section{Solution via double-obstacle HJI VI and SPP\label{sec:solution}}
One direct way of solving the problem formulated in Section \ref{sec:formulation} is by solving a single-obstacle HJI VI \cite{mitchell05,bokanowski10, Margellos11, Margellos13}. In this approach, one considers the joint time-invariant dynamics of the entire system, $f(\x,\ctrl)$, and defines the static goal set and the static avoid set in the joint state space of all vehicles. The goal set encodes the joint states representing all vehicles being at their target sets, and the avoid set encodes the joint states representing all unsafe configurations. These sets are defined as sub-zero level sets of appropriate implicit surface functions $\setf(\x)$ where $\x\in\set \Leftrightarrow \setf(\x)\le 0$. Having defined the implicit surface functions, the HJI VI (\ref{eq:HJIPDE}) is then solved backwards in time with the implicit surface function representing the terminal set $\goalf(\x)$ as the initial condition and the implicit surface function representing the avoid set $\avoidf(\x)$ as an effective constraint:

\begin{equation}
\begin{aligned}
	\label{eq:HJIPDE}
	\max\big\{D_t\soln + \min \left[0, H\left(\x,D_{\x}\soln\right)\right], -\avoidf(\x)-\soln(\x,t) \big\}= 0,\\
\soln(\xj,0) = \goalf(\x)	 
\end{aligned}
\end{equation}
\noindent with the optimal Hamiltonian $H\left(\xj,p\right) = \min_{\ctrl \in \ctrlin} p \cdot f(\x,\ctrl).$

The solution $\soln(\x, t)$ is the implicit surface function representing the reach-avoid set $\RA(t)$, which defines the set of states from which the system has a control to drive the state at time $t$ to the goal set $\goal$ at time $0$ while staying out of the avoid set $\avoid$ at all times. Note that the joint dynamics, goal set, and avoid set must be time-invariant. Time-varying dynamics and sets can be treated by augmenting the state space with time as an auxiliary state \cite{bokanowski11}; however, this state augmentation comes at a large computational expense.

The direct solution described above has been successfully used to solve a number of problems involving up to a pair of vehicles \cite{mitchell05, ding08, huang11, chen14}. However, since numerical methods for solving a PDE or a VI involve gridding up the state space, the computation complexity scales exponentially with the number of dimensions in the joint state. This makes the single-obstacle HJI VI inapplicable for problems involving three or more vehicles. Therefore, instead of solving a single-obstacle HJI VI in the joint state space in $\R^n=\R^{\sum_i n_i}$, we will consider the problem in in $\R^{n_i}$ and solve a sequence of \textit{double-obstacle} HJI VIs introduced in \cite{fisac15}. By doing so, we take advantage of the fact that time-varying targets, obstacles, and dynamics can be handled by the double-obstacle HJI VIs (but not by the single-obstacle HJI VI without incurring significant computational expense), making the analysis of the problem tractable. Furthermore, even if the dimensionality of the problem is sufficiently low for computing a numerical solution to the single-obstacle HJI VI, its inability to handle time-varying systems would still limit us to only consider problems in which the required time of arrival is common across all vehicles: $\tf_i = \tf \;\forall i$.

We first describe the framework for computing reach-avoid sets with arbitrary terrain, domain, moving obstacles, and moving target sets based on \cite{fisac15}. As with the single-obstacle HJI VI, sets are defined as sub-zero level sets of implicit surface functions; however, crucially, these implicit surface functions can be time-varying in the double-obstacle HJI VI without increasing computational complexity. Being able to compute reach-avoid sets with moving obstacles allows us to overcome the computational intractability described above by sequentially performing path planning for one vehicle at a time in order of priority, while treating higher-priority vehicles as moving obstacles. The target set is defined in the same way as in the single-obstacle HJI VI; the avoid set is by convention defined as the complement of the state constraint set in the double-obstacle HJI VI.

\subsection{Reachability via HJI VI}
We first state the result given in \cite{fisac15}, and then specialize the result to the problem formulation given in Section \ref{sec:formulation}. Consider a general nonlinear system describing the state evolution of two players in a differential game for $t\in[0,T]$.

\begin{equation}
\dot{x}(t) = f(t, x, u, d), \quad x(0) = x
\end{equation}

\noindent where $x$ is the joint state, $u$ is the control input for player 1, and $d$ is the control input for player 2. Their joint dynamics $f$ is assumed to be bounded, Lipschitz continuous in $x$ for any fixed $u,d$ and $t$, and measurable in $t,u,d$ for each $x$. Given control functions $u(\cdot), d(\cdot)$, there exists a unique trajectory $\phi_x^{u,d}((\tau),\tau)$ \cite{coddington55}. Player 1 wishes to minimize, and player 2 wishes to maximize the following cost functional:

\begin{equation}
\begin{aligned}
&\mathcal{V}\big(t,x,u(\cdot),d(\cdot)\big) \\
&\quad = \min_{\tau\in[t,T]}\max\big\{l(\phi_{x(0)}^{u,d}(\tau),\tau),\max_{s\in[t,\tau]} g(\phi_{x(0)}^{u,d}(s),s)\big\}
\end{aligned}
\end{equation}

The value of the game is thus given by

\begin{equation}
\begin{aligned}
\soln(x,t):=\sup_{\delta[u](\cdot)}\inf_{u(\cdot)}\mathcal{V}\big(t,x,u(\cdot),\delta[u](\cdot)\big)
\end{aligned}
\end{equation}

\noindent where player 2 chooses a nonanticipative strategy $d(\cdot) = \delta[u](\cdot)$, under which the control signal $d(t)$ is chosen in response to player 1's control function up to time $t$, $u(\tau),\tau\le t$ \cite{mitchell-thesis}. The value of the game characterizes reach-avoid set, or all the states from which player 1 can reach the target $\goal$ encoded by the implicit surface function $\goalf(x,t)$, while staying within some state constraint set $\constr$ encoded by the implicit surface function $\constrf(x,t)$, despite the adversarial actions of player 2. The value function is the unique viscosity solution \cite{crandall84} to the following single-obstacle HJI VI \cite{fisac15}:

\begin{equation}
\label{eq:HJIVI_full}
\begin{aligned}
\max\Big\{&\min\big\{D_t V + H\left(x, D_x V,t\right),l(x,t)-V(x, t)\big\}\\
& g(x,t)-V(x,t)\Big\}=0, \quad t\in[0,T], \quad x\in\R^n\\
&V(x,T) = \max\big\{l(x,T),g(x,T)\big\},  \quad x\in\R^n
\end{aligned}
\end{equation}

The proof is given in \cite{fisac15} and is based on viscosity solution theory \cite{evans84, barron90}.

Now consider the system with dynamics given by (\ref{eq:dyn}). Given a time-varying target set $\target_i(t)$ and obstacle $\avoid_i(t)$ that vehicle $i$ must avoid, we define implicit surface functions $\goalf(\x_i,t), \constrf(\x_i,t)$ such that $\x_i\in\target_i(t)\Leftrightarrow \goalf_i(\x_i,t)\le 0,\x_i\notin \avoid_i(t) \Leftrightarrow \constrf_i(x,t)\le 0$. Now, the problem formulated in Section \ref{sec:formulation} becomes one in which vehicle $i$ chooses a control function $\ctrl_i(\cdot)$ to minimize the following cost functional:

\begin{equation}
\label{eq:cost}
\begin{aligned}
&\mathcal{V}_i\big(t,\x_i, \ctrl_i(\cdot)\big) \\
&\qquad= \min_{\tau\in[t,T]}\max\big\{\goalf_i(\x_i(\tau),\tau),\max_{s\in[t,\tau]} \constrf_i(\x_i(s),s)\big\}
\end{aligned}
\end{equation}

Note here, we have an optimal control problem involving only one vehicle and no adversary (given $\constrf_i(\x_i(s),s)$), unlike in the case of the HJI VI (\ref{eq:HJIVI_full}). Now, specializing (\ref{eq:HJIVI_full}) to our optimal control problem, the value function that characterizes the reach-avoid set $\RA_i(t)$ is $\soln_i(\x_i, t)$, where $\x_i \in \RA_i(t) \Leftrightarrow \soln_i(\x_i, t) \le 0$. $\soln_i(\x_i, t)$ is the viscosity solution \cite{crandall84} of the HJI VI

\begin{equation}
\label{eq:HJI}
\begin{aligned}
\max\big\{\min\{D_t \soln_i + H_i\left(\x_i, D_{\x_i} \soln_i,t\right),\goalf_i(\x_i,t)-\soln_i(\x_i, t)\}\\
\quad \constrf_i(\x_i,t)-\soln_i(\x_i,t)\big\}=0, t\in[\tnow_i,\tf_i], \x_i\in\R^{n_i}\\
\soln_i(\x_i,\tf_i) = \max\left\{\goalf_i(\x_i,\tf_i),\constrf_i(\x_i,\tf_i)\right\}, \x_i\in\R^{n_i}
\end{aligned}
\end{equation}

\noindent where the Hamiltonian $H_i(t, \x_i, p)$ and optimal control $\ctrl_i$ are given by

\begin{equation}
\begin{aligned}
H_i(t,\x_i,p) &= \min_{\ctrl_i\in\ctrlin_i} p \cdot f_i(t,\x_i,\ctrl_i) \\
\ctrl^*_i &= \arg \min_{\ctrl_i} H_i(t,\x_i,p)
\end{aligned}
\end{equation}

\subsection{Sequential Path Planning}
In order to use (\ref{eq:HJI}) to perform SPP, we first define the moving obstacles induced by higher-priority vehicles. Specifically, for vehicle $i$, we define the moving obstacles $\mobs^i_j(t)$ induced by vehicles $j=1,\ldots,i-1$, given their known trajectories $\x_j (\cdot)$, to be $\mobs^i_j(t) := \{\x_i: \pos_i \in \danger_{ij}(\x_j(t))$.

Each vehicle $i$ must avoid being in $\mobs^i_j(t)$ for each $j=1,\ldots,i-1$ and for all time $t$, as well as avoid being in static obstacles $\obs$ in the domain. Therefore, for the \ith vehicle, we compute the reach-avoid set with the following time-varying avoid set $\avoid_i(t)$ and goal set $\goal_i(t)$:

\bq
\begin{aligned}
\avoid_i(t) &:= \{\x_i: \pos_i \in \obs\} \cup \Big(\bigcup_{j=1,\ldots,i-1} \mobs^i_j(t)\Big)\\
\goal_i(t) &:= \target_i, t\le \tf_i
\end{aligned}
\eq

The goal set is represented by the implicit surface function $\goalf_i(\x,t)$, where $\goalf_i(\x_i,t)\le0\Leftrightarrow \x_i(t)\in \goal_i(t)$. The state constraint set in the HJI VI is defined as the complement of the avoid set, $\avoid_i^c(t)$, and is represented by the implicit surface function $\constrf(\x_i,t)$, where $\constrf(\x_i,t)\le0 \Leftrightarrow \x_i\notin \avoid_i(t)$. For both $\goalf_i(\x_i,t)$ and $\constrf(\x_i,t)$, we use the signed distance function (in $\x_i$) to the sets $\goal_i(t)$ and $\avoid_i^c(t)$, respectively.

Now, we can solve the double-obstacle HJI VI (\ref{eq:HJI}). The solution $\soln(\x_i,t)$ represents the reach-avoid set $\RA(t)$: $\soln(\x_i,t)\le0\Leftrightarrow \x_i(t)\in\RA(t)$. $\RA(t)$ is the set of states at starting time $t$ from which vehicle $i$ can arrive at $\target_i$ at or before time $\tf_i$ while avoiding obstacles and danger zones of all higher-priority vehicles $j=1,\ldots,i-1$. 

Alternatively, given an initial state $\x_i^0$, we can solve (\ref{eq:HJI}) to some $\ti_i = \inf\{t:\x_i^0 \in \RA(t)\}$. This represents the latest time that vehicle $i$ must depart from its initial position in order to reach $\target_i$ while avoiding obstacles and all danger zones of higher-priority vehicles $j=1,\ldots,i-1$.

The optimal control is given by

\bq
\label{eq:ctrl_syn}
\ctrl_i(t) = \arg \min \ham_i \left(t, D_{\x_i} \soln(\x_i, t), \soln(\x_i, t)\right)
\eq

Observe that since each vehicle $i$ is guaranteed to be safe with respect to higher priority vehicles $j=1,\ldots,i-1$, the safety of all vehicles, including lower-priority vehicles, can also be guaranteed.

\section{Results: Four Vehicles with Constrained Turn Rate}
Consider four vehicles with states $\x_i = [x_i, y_i, \theta_i]^\top$ modeled using a horizontal kinematics model with the following dynamics for $t \in[\tnow_i, \tf_i],i=1,2,3,4$:

\begin{equation}
\begin{aligned}
\dot{x}_i &= v_i \cos(\theta_i) &\\
\dot{y}_i &= v_i \sin(\theta_i) &\qquad \x_i(\tnow_i) = \x_i^0 \\
\dot{\theta}_i &= \omega_i &\qquad|\omega_i| \le \bar{\omega}_i 
\end{aligned}
\end{equation}

\noindent where $(x_i, y_i)$ is the position of vehicle $i$, $\theta_i$ is the heading of vehicle $i$, and $v_i$ is the speed of vehicle $i$. The control input $\ctrl_i$ of vehicle $i$ is the turning rate $\omega_i$, whose absolute value is bounded by $\bar{\omega}_i$. For illustration, we chose $\bar{\omega}_i=1 \forall i$ and assume $v_i=1$ is constant; however, our method can easily handle the case in which $\bar{\omega}_i$ differ across vehicles and $v_i$ is a control input. Optimizing the Hamiltonian associated with vehicle $i$, $\ham_i(t, D_{\x_i}\soln_i(\x_i,t), \soln_i(\x_i,t))$, we can obtain the optimal control
%
%
\bq
\omega_i(t) = -\bar{\omega}_i\frac{D_{\theta_i}\soln_i(\x_i,t)}{\left| D_{\theta_i}\soln_i(\x_i,t) \right|}
\eq

The vehicles have initial conditions and STA as follows:
\bq
\begin{aligned}
\x_1^0 &= (-0.5, 0, 0), &\tf_1 &= 0\\
\x_2^0 &= (0.5, 0, \pi), &\tf_2 &= 0.2\\
\x_3^0 &= \left(-0.6, 0.6, 7\pi/4\right), &\tf_3 &= 0.4\\
\x_4^0 &= \left(0.6, 0.6, 5\pi/4\right), &\tf_4 &= 0.6\\
\end{aligned}
\eq

\begin{figure}
	\centering
	\includegraphics[width=0.35\textwidth]{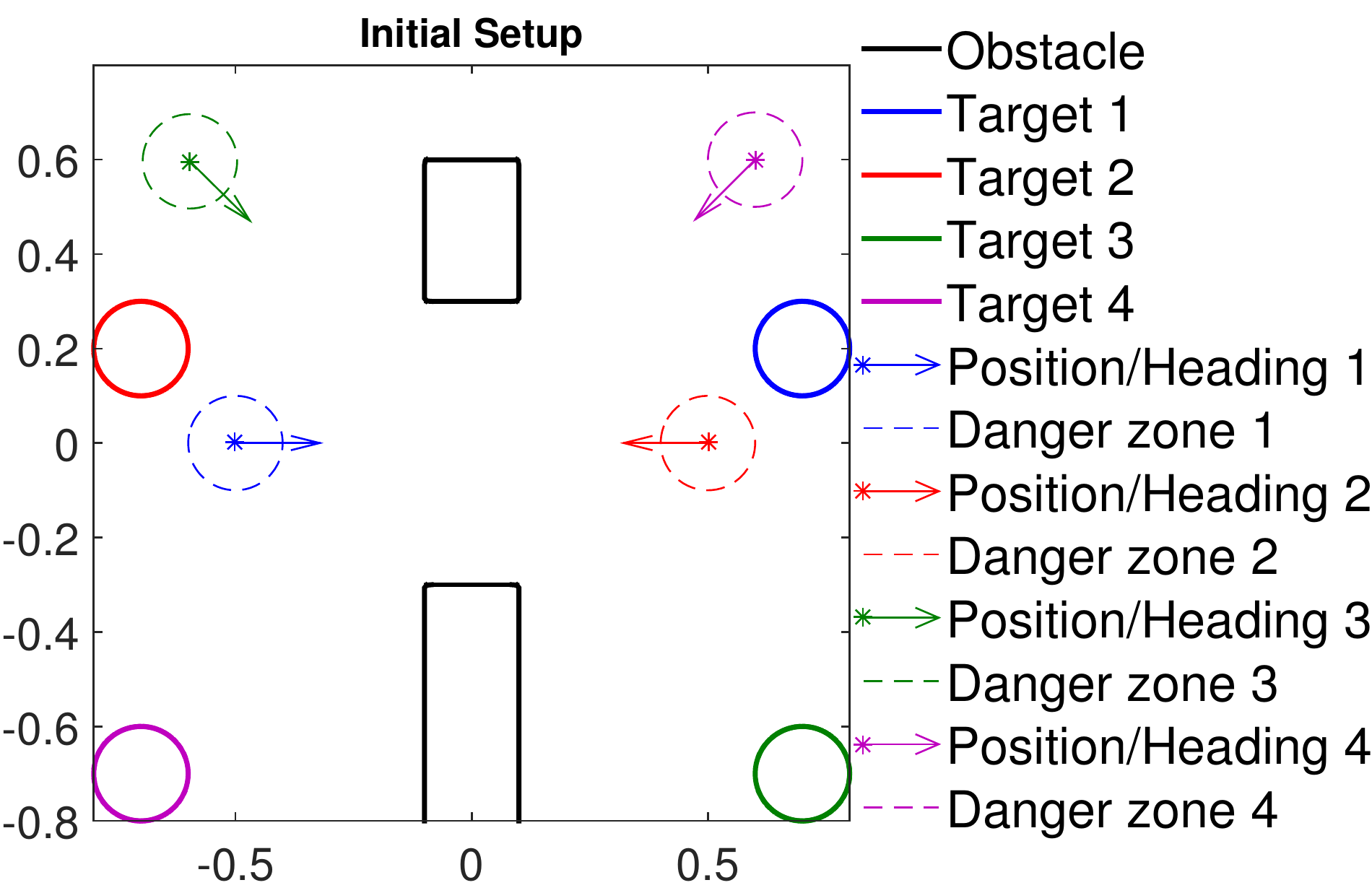}
	\caption{Initial configuration of the four-vehicle example.}
	\label{fig:dubins_ic}
\end{figure}

The target sets $\target_i$ of the vehicles are all 4 circles of radius $0.1$ in the domain. The centers of the target sets are at $(0.7, 0.2), (-0.7, 0.2), (0.7, -0.7), (-0.7, -0.7)$ for vehicles $i=1,2,3,4$, respectively. The obstacles are rectangles near the middle of the domain. 
The setup for this example is shown in Figure \ref{fig:dubins_ic}. 

The joint state space of this system is twelve-dimensional, intractable for analysis using the single-obstacle HJI VI (\ref{eq:HJIPDE}). Therefore, we will repeatedly solve the double-obstacle HJI VI (\ref{eq:HJI}) to compute the reach-avoid sets from targets $\target_i$ for vehicles $1,2,3,4$, in that order, with moving obstacles induced by vehicles $j=1,\ldots,i-1$. We will also obtain $\ti_i,i=1,2,3,4$, the LSTs for each vehicle in order to reach $\target_i$ by $\tf_i$.

\begin{figure}
	\centering
	\includegraphics[width=0.45\textwidth]{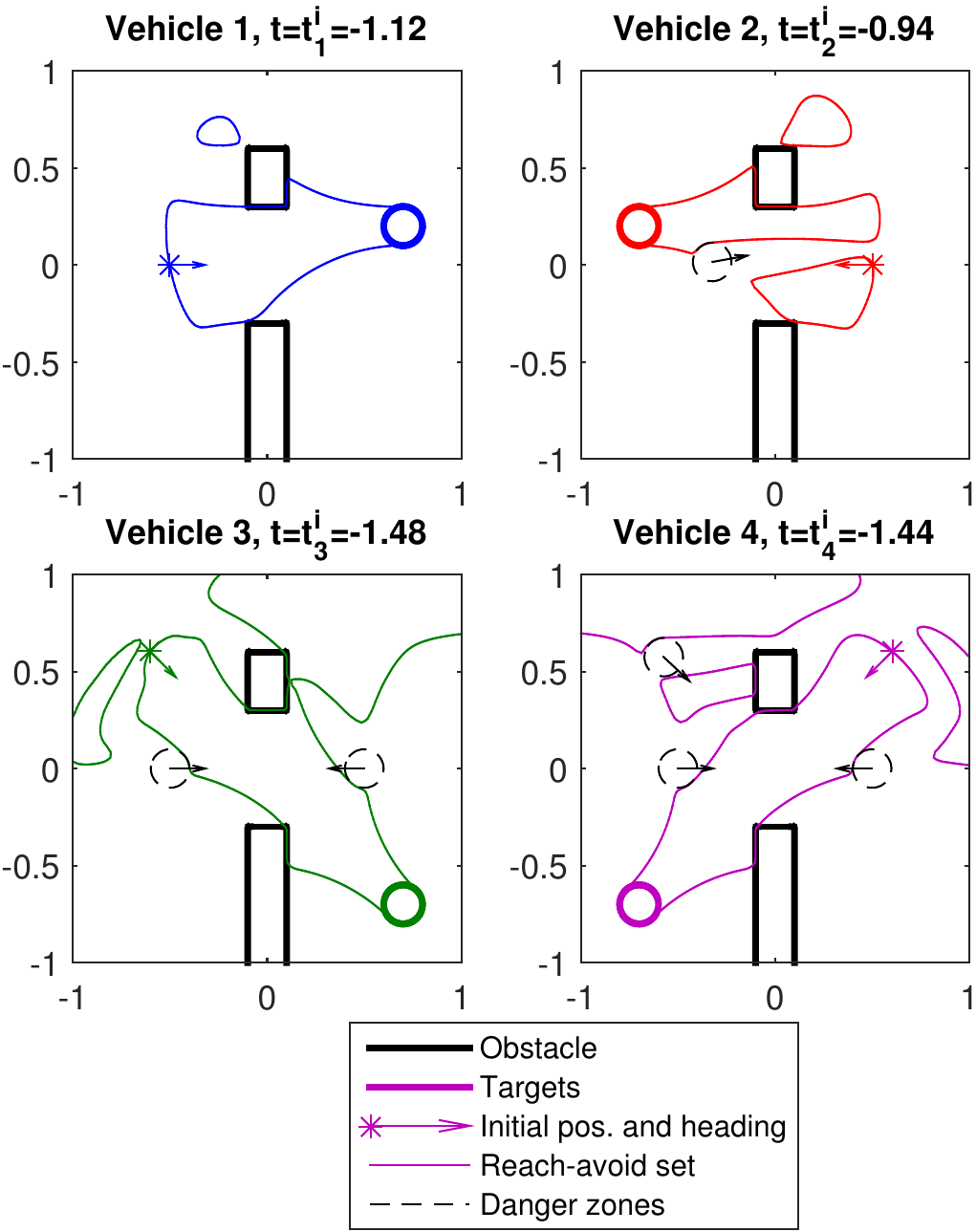}
	\caption{Reach-avoid sets at $t=\ti_i$ for vehicles $1,2,3,4$, sliced at initial headings $\theta_i^0$. Black arrows indicate direction of obstacle motion. Due to the turn rate constraint, the presence of static obstacles $\obs$ and time-varying obstacles induced by higher-priority vehicles $\mobs^i_j(t)$ carves ``channels" in the reach-avoid set, dividing it up into multiple ``islands".}
	\label{fig:dubins_reach_all}
\end{figure}

\begin{figure}
	\centering
	\includegraphics[width=0.45\textwidth]{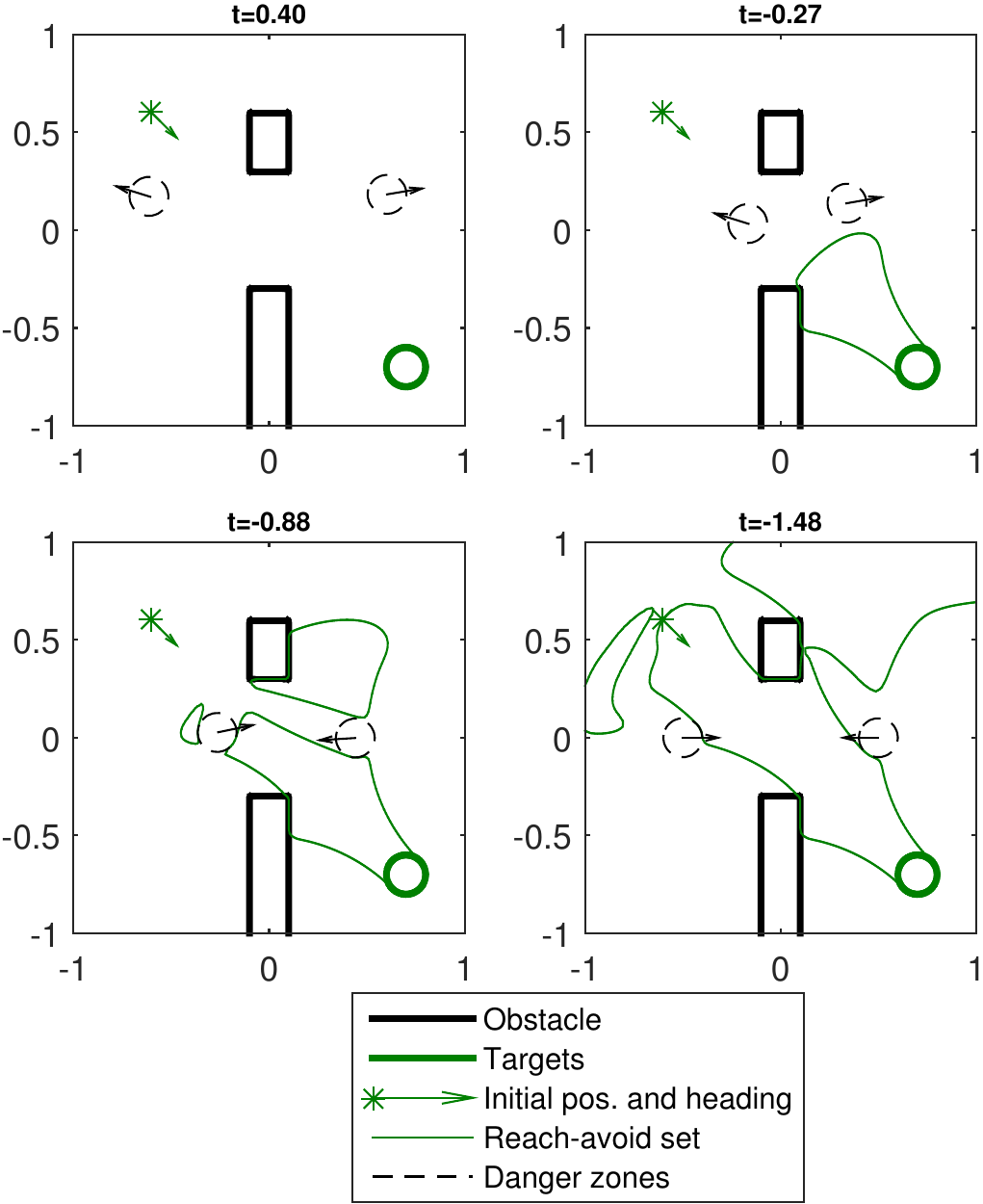}
	\caption{Time evolution of the reach-avoid set for vehicle $3$, sliced at its initial heading $\theta_3^0=\frac{7\pi}{4}$. Black arrows indicate direction of obstacle motion. Initially, the reach-avoid set grows unobstructed by obstacles, as shown in the top subplots. Then, in the bottom subplots, the static obstacles $\obs$ and the induced obstacles of vehicles $1$ and $2$, $\mobs^3_1,\mobs^3_2$, carve out ``channels" in the reach-avoid set.}
	\label{fig:dubins_reach_3}
\end{figure}

\begin{figure}
	\centering
	\includegraphics[width=0.45\textwidth]{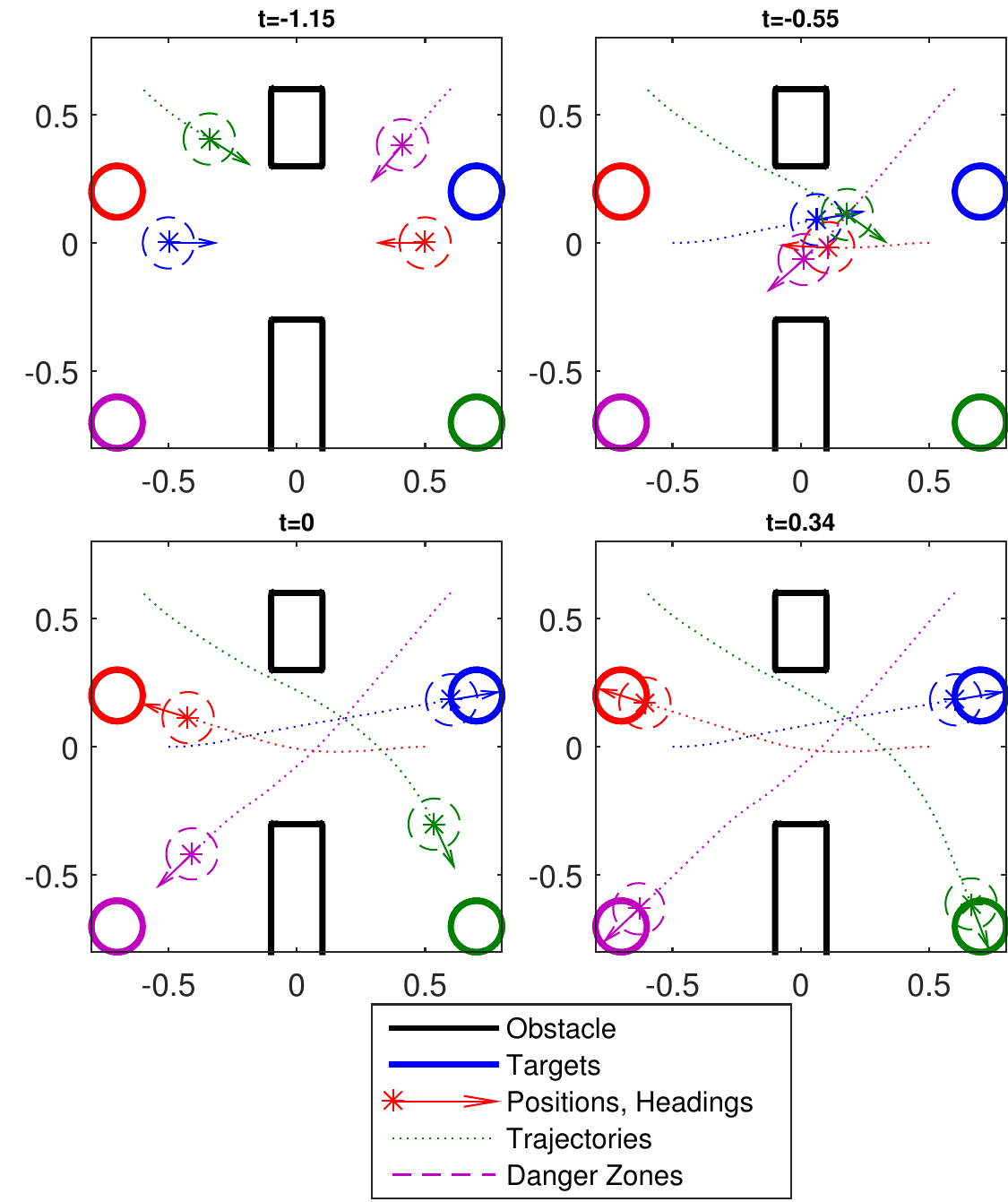}
	\caption{The planned trajectories of the four vehicles. In the left top subplot, only vehicles $3$ (green) and $4$ (purple) have started moving, showing $\ti_i$ is not common across the vehicles. Right top subplot: all vehicles have come within very close proximity, but none is in the danger zone another. Left bottom subplot: vehicle $1$ (blue) arrives at $\target_1$ at $t=0$. Right bottom subplot: all vehicles have reached their destination, some ahead of the STA $\tf_i$.}
	\label{fig:dubins_result}
\end{figure}

Figures \ref{fig:dubins_reach_all}, \ref{fig:dubins_reach_3}, and \ref{fig:dubins_result} show the results. Since the state space of each vehicle is 3D, the reach-avoid set is also 3D. To visualize the results, we slice the reach-avoid sets at the initial heading angles $\theta_i^0$. Figure \ref{fig:dubins_reach_all} shows the 2D reach-avoid set slices for each vehicle at its LSTs $\ti_1=-1.12, \ti_2=-0.94,\ti_3=-1.48,\ti_4=-1.44$ determined from our method. The obstacles in the domain $\obs$ and the obstacles induced by other vehicles inhibit the evolution of the reach-avoid sets, carving out thin ``channels" that separate the reach-avoid set into different ``islands". One can see how these channels and islands form by examining the time evolution of the reach-avoid set, shown in Figure \ref{fig:dubins_reach_3} for vehicle 3. 

Finally, Figure \ref{fig:dubins_result} shows the resulting trajectories of the four vehicles. The subplot labeled $t=-0.55$ shows all four vehicles in close proximity without collision: each vehicle is outside of the danger zone of all other vehicles. The actual arrival times of vehicles $i=1,2,3,4$ are $0, 0.19, 0.34, 0.31$, respectively. It is interesting to note that for some vehicles, the actual arrival times are earlier than the STAs $\tf_i, i=1,2,3,4$. This is because in order to arrive at the target by $\tf_i$, these vehicles must depart early enough to avoid major delays  resulting from the induced obstacles of other vehicles; these delays would have lead to a late arrival if vehicle $i$ departed after $\ti_i$.

\section{Conclusion}
We have presented a problem formulation that allows us to consider the multi-vehicle trajectory planning problem in a tractable way by planning trajectories for vehicles in order of priority. In order to do this, we modeled higher-priority vehicles as time-varying obstacles. We then solved a double-obstacle HJI VI to obtain the reach-avoid set for each vehicle. The reach-avoid set characterizes the region from which each vehicle is guaranteed to arrive at its target within a time horizon, while avoiding collision with obstacles and higher-priority vehicles. The solution also gives each vehicle a latest start time as well as the optimal control which guarantees that each vehicle safely reaches its target on time. 


\bibliographystyle{IEEEtran}
\bibliography{references}
\end{document}